\documentclass[elsart]{revtex4}

\usepackage{graphicx}

\begin{document}
%\twocolumn[\hsize\textwidth\columnwidth\hsize\csname @twocolumnfalse\endcsname

\draft

\title{Barrier formation at metal/organic interfaces: dipole formation and the
Charge Neutrality Level}

\author{H. V\'azquez$^{\dagger}$, F. Flores$^{\dagger}$, R.
Oszwaldowski$^{\dagger,\ddagger}$, 
J. Ortega$^{\dagger}$, R. P\'erez$^{\dagger}$ and A. Kahn$^{\P}$}

\address{ $^{\dagger}$
Departamento de F\'{\i}sica Te\'orica de la Materia Condensada,
Universidad Aut\'onoma de Madrid, E-28049 Madrid, Spain. \\
$^{\ddagger}$ Instytut Fizyki Miko{\l}aja Kopernika, Grudzi\c{a}dzka 5, 87-100 
Toru\'n, Poland. \\
$^{\P}$ Department of Electrical Engineering, Princeton University,
Princeton, NJ 08544, USA. }

%\date{\today}

\begin{abstract}

The barrier formation for metal/organic semiconductor interfaces is analyzed within the
Induced Density of Interface States (IDIS) model. Using weak chemisorption
theory, we calculate the induced density of states in the organic energy gap and
show that it is high enough to control the barrier formation. We calculate the
Charge Neutrality Levels of several organic molecules (PTCDA, PTCBI and
CBP) and the interface Fermi level for their contact with a Au(111) surface.
We find an excellent agreement with the experimental evidence and conclude that
the barrier formation is due to the charge transfer between the metal and the
states induced in the organic energy gap.

\end{abstract}

\maketitle

%\pacs{PACS numbers: 
%}

%\bigskip

%\narrowtext

\section{Introduction}

The field of electronic materials based on molecular films is developing very
fast. Designing new organic-based materials and devices requires a detailed 
knowledge of basic processes, such as those controlling the formation of 
metal/organic interface barriers \cite{Ishii_book}.

The evolution of our understanding of this aspect of organic interfaces 
follows a path which reminds us of the slow process of understanding
inorganic semiconductor/metal interfaces. Twenty-five years ago, the main
problem was to understand the mechanism of the formation of
the Schottky barrier. Starting with the Schottky and Bardeen models,
research developed new ideas based on the Defect model \cite{Defect} and the Induced Density of
Interface States model \cite{IDIS}. The present consensus on
inorganic/metal interfaces is that, unless the interface has
many defects, the Schottky barrier formation is controlled by ``intrinsic"
interface states induced by the interaction between the inorganic semiconductor
and the metal.

Several models have been advanced to explain organic semiconductor/metal interface barrier 
formation. The Schottky-Mott model was originally believed to hold for these
interfaces, assuming that no interface dipole is formed at the junction, a
situation which was subsequently disproved in most cases \cite{SM, Narioka}. 
At reactive interfaces, the 
metal-molecule chemical reaction creates gap
states that pin the Fermi level, a situation that is analagous to that described
by the Unified Defect model proposed for inorganic semiconductor/metal
interfaces \cite{Schen}. Compression of the metal surface electronic tail by the organic
molecules, leading to a change in the metal workfunction, has also been
suggested as a mechanism operating in these interfaces \cite{IshiiAdvM}.

\begin{figure}[htbp]
\begin{center}
\vspace{0.50cm}
\includegraphics[width=7cm, angle=0] {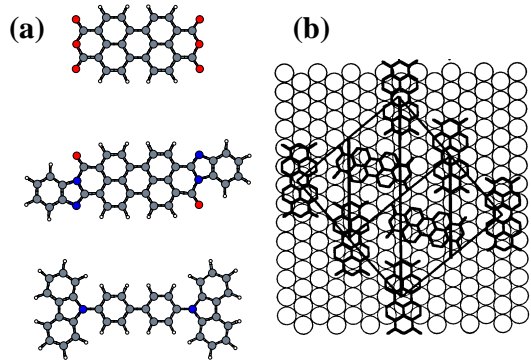}
\end{center}
\caption{(a) Organic molecules studied in this paper: PTCDA, PTCBI and CBP; (b)
Two-dimensional pattern of a PTCDA crystal (from \cite{refPTCDAcrystal}).
}
\label{Fig1}
\end{figure}

This communication focuses on non-reactive interfaces between metals and thin 
films of low weight organic molecules. We study several organic 
semiconductor/metal interfaces within the IDIS model \cite{IDIS} and show that 
the induced densities of states at these junctions are large enough to
control the interface barrier formation. We analyze 
the chemical interaction between Au and several 
organic molecules: 3, 4, 9, 10- perylenetetracarboxylic
dianhydride (PTCDA), 3, 4, 9, 10- perylenetetracarboxylic
bisbenzimidazole (PTCBI) and 4, 4', -N, N'- dicarbazolyl biphenyl (CBP)
(Figure~\ref{Fig1}(a)). Our quantum-mechanical
analysis shows how the weak chemical interaction creates, nevertheless, a local
density of states in the organic energy gap, which is large enough to make the
IDIS model applicable and the definition of a Charge Neutrality Level (CNL) for 
the organic molecule meaningful. Our results for these junctions explain their 
pinning behaviour, which is characterized by the interface slope parameter,
$S = \frac {d E_{F}}{d \phi_{M}}$. A low value of $S$ (as is the case of 
PTCDA/Au and PTCBI/Au) corresponds to strong Fermi level pinning, whereas a 
higher value of $S$ (CBP/Au is an example) means that the change of the 
barrier height with the metal workfunction is larger: as will be discussed 
below, this is associated 
with a smaller density of states induced in the organic energy gap. 

\section{Model and Theoretical solution}

Consider, initially, the case of a PTCDA/Au interface. Experimental evidence
indicates that PTCDA molecules lie flat on the Au surface, and that the 
PTCDA monolayer has the
two-dimensional pattern shown in Figure~\ref{Fig1}(b). PTCDA crystals can be
thought of being formed by the repetition of this layer along the direction
perpendicular to the surface \cite{ForrestRev}. It is important to realize
that, as is typical of organic crystals, intermolecular bonds are weak van der 
Waals bonds that preserve the individuality of the molecules.
This simplifies the analysis of the PTCDA-Au interaction, reducing it
to the case of a single molecule deposited with its plane parallel to the 
surface.

We should stress that, in principle, the molecule-molecule interaction induces 
some (small) broadening of the electronic levels of each individual molecule,
but does not create an electron density of states in the molecular energy gap. Since the
Schottky barrier formation depends on these gap states, we can safely neglect
the molecule-molecule interaction and consider only the single molecule-metal
interaction.

Our analysis of the organic semiconductor/metal interface is made in several steps. First,
the organic molecule is analyzed using a DFT-LCAO method \cite{Pou}; this is a DFT-based
theory, which uses a local-orbital basis and the orbital occupation for
describing the exchange and correlation energies of the system. It has been
shown elsewhere that this approach is equivalent to other more conventional DFT
methods that use  a local exchange and correlation energy \cite{Rafal}.

The main problem with DFT methods for organic molecules (and other small
molecules) is that their one-electron eigenvalues do not represent real electron
or hole excitations. In particular, the DFT energy gap is not directly related
to either the transport or the optical gap. The advantage of our approach is that,
by using a variation of Koopman's theorem \cite{Rafal}, we can calculate the molecular 
electronic levels by introducing a relaxation energy in the DFT method that is
directly related to the exchange-correlation energy. We have shown for
PTCDA \cite{PTCDA_CNL}
that, although this effect is important, the relaxation of the
molecular orbitals calculated within DFT is negligible \cite{Rafal}. This means that, 
after the DFT-LCAO calculation for the organic molecule, we need to renormalize 
the energy levels by means of Koopman's relaxation energies in order to obtain 
a realistic
energy spectrum. We can keep, however, the electronic wavefunctions of the
molecule obtained from the DFT calculation.

\begin{figure}[htbp]
\begin{center}
\vspace{0.50cm}
\includegraphics[width=7cm, angle=0] {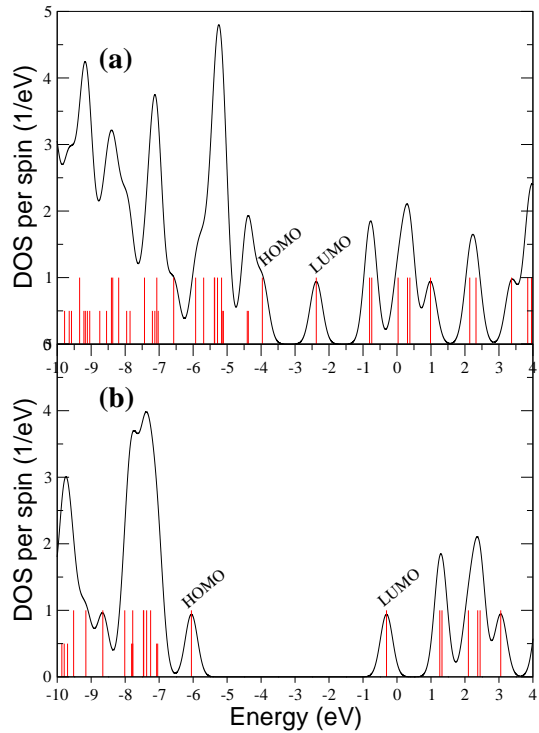}
\end{center}
\caption{DFT-LCAO energy spectrum (long bars: $\pi$ states; short bars: $\sigma$
states) before (a) and after (b) Koopman's relaxation energies; 
the relaxation in the wavefunctions, however, is small. The DOS is obtained by
introducing a 0.5 eV FWHM Gaussian broadening for each level.
}
\label{Fig2}
\end{figure}

Figure \ref{Fig2} illustrates this situation for PTCDA. Panel (a) shows the energy
spectrum calculated within DFT-LCAO, while panel (b) shows the spectrum after
introducing Koopman's relaxation energies. Note how the PTCDA energy gap
increases from 1.6 eV to 5.7 eV due to these relaxation effects; the molecular
wavefunctions associated with the ionization (HOMO) and affinity (LUMO) levels 
are assumed to
be practically the same before and after the energy relaxation.

There are also solid-state effects associated with long-range electronic polarization.
These are mainly associated with screening of the metal and the organic crystal 
of the electronic field created by the extra charge (electron or hole) introduced
in the molecule \cite{Tsiper}. This correction reduces the energy gap by $\sim$1.5eV, although the
ionization and affinity wavefunctions of the molecule are not expected to present
important modifications. Other polarization effects, due to lattice relaxation and
vibronic effects only introduce small corrections, around 0.2eV, further reducing
the molecular energy gap \cite{Schen}. Regarding PTCDA, we fit the molecular transport gap to
3.2eV, which presumably takes into account all the effects discussed above \cite{Schen}; this is
done by introducing a rigid shift between the empty and occupied states of
Figure \ref{Fig2}(b). For PTCBI and CBP, we fit the transport gaps to 3.1 eV and 5.1 eV
respectively, values which were deduced from experimental optical
gaps and exciton binding energies \cite{OptGaps}.

In a second step, we calculate the induced density of states at the 
organic semiconductor/metal
interface using chemisorption theory in the limit of weak interaction between
the two systems \cite{Zangwill}.  In our model, we assume the organic molecule (PTCDA,
PTCBI or CBP) to be deposited flat on the metal surface, at a distance $d$ from the 
last metal layer, which we will take
for the rest of this paper to be Au(111).

In our analysis, we start with the organic molecule wavefunctions,
$\psi_{i}$, obtained from the DFT-LCAO method discussed above, and the metal
density of states matrix, $\rho_{\alpha \beta}(E)$, where $\alpha$ and $\beta$ 
refer to the local-orbital basis used to describe the metal
properties, which are calculated using the DFT local-orbital code 
$Fireball$ \cite{Fireball}. In 
the limit of weak PTCDA-metal interaction, the main effect of the metal is to
broaden the molecular levels $E_{i}$ by the quantity $\Gamma_{i}$ \cite{Zangwill}:

\begin{equation}
\Gamma_{i} = 2 \pi \sum_{\nu} \; |T_{i\nu}|^{2} \; \delta (E_{\nu} - E_{i}), \label{gammai}
\end{equation}

where $T_{i\nu}$ is the hopping  interaction between the molecular orbital 
$\psi_{i}$ and the metal eigenfunction, $\psi_{\nu}$. Equation \ref{gammai} can
be rewritten in a more convenient way by using the local-orbital basis for the
molecule and the metal. Writing $\psi_{i} = \sum_{j} c_{ij} \phi_{j}$, 
equation \ref{gammai} takes the form

\begin{equation}
\Gamma_{i} = 2 \pi \; \sum_{j j^{\prime} \alpha \beta} c_{ij} \; T_{j\alpha} \; 
\rho_{\alpha\beta} (E_{i}) \; T_{\beta j ^{\prime}} \; c_{j ^{\prime} i}.
\label{gammaicij}
\end{equation}

Neglecting off-diagonal terms with $j \neq j^{\prime}$ and $\alpha \neq 
\beta$, which tend to cancel each other out, equation \ref{gammaicij} is
further simplified to

\begin{equation}
\Gamma_{i} = 2 \pi \; \sum_{j, \alpha} |c_{i,j}|^{2}  |T_{j,\alpha}|^{2}
\rho_{\alpha,\alpha} (E_{i}).
\label{gammai6s}
\end{equation}

\begin{figure}[htbp]
\begin{center}
\vspace{0.50cm}
\includegraphics[width=7cm, angle=0] {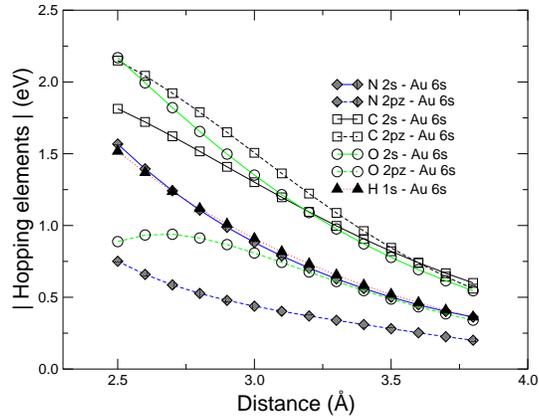}
\end{center}
\caption{Hopping elements between the atoms which make up the organic molecule
and Au as a function of distance.
}
\label{Fig3}
\end{figure}

In our calculation, we have only included the interaction of the Au 6s orbital
with the different orbitals of the organic molecules, C $2s$ and $2p$, N $2s$
and $2p$, O $2s$ and $2p$, and H $1s$. This implies that in equation
\ref{gammai6s}, $\alpha$ refers only to the Au 6s orbitals. Figure \ref{Fig3} shows
these interactions as a function of the Au-atom distance, which obviously
depends on the organic-metal distance, $d$. 

The organic-metal separation is a difficult issue. First, no experimental value 
exists for these systems. Second, conventional DFT codes cannot be used to
calculate $d$, due to the weak van der Waals interaction between the metal and 
the organic molecule \cite{ForrestRev}. Using indirect information based on: (a) the PTCDA-PTCDA
stacking distance, $d \sim 3.2$ {\AA}, and, (b) the atomic radius of Au 
($\sim 0.5$ {\AA} larger than that of C), we assume the
distances between the last Au layer and the plane of the organic molecules
considered in this paper to be around $3.5 \pm 0.3$ {\AA}. We make use of this
value to calculate $\Gamma_{i}$ from equation \ref{gammai6s}.

Once $\Gamma_{i}$ is calculated, each molecular level $E_{i}$ contributes to
the organic LDOS with the Lorentzian function 

\begin{equation}
\frac {1} {\pi} \frac {\Gamma_{i}/2} {(E-E_{i})^{2} + {(\frac{\Gamma_{i}}
{2})^{2}}}.
\end{equation}

The total LDOS is obtained by adding all the contributions from the
different molecular orbitals.

\section{Results and Conclusions}\label{sec:results}

\begin{figure}[htbp]
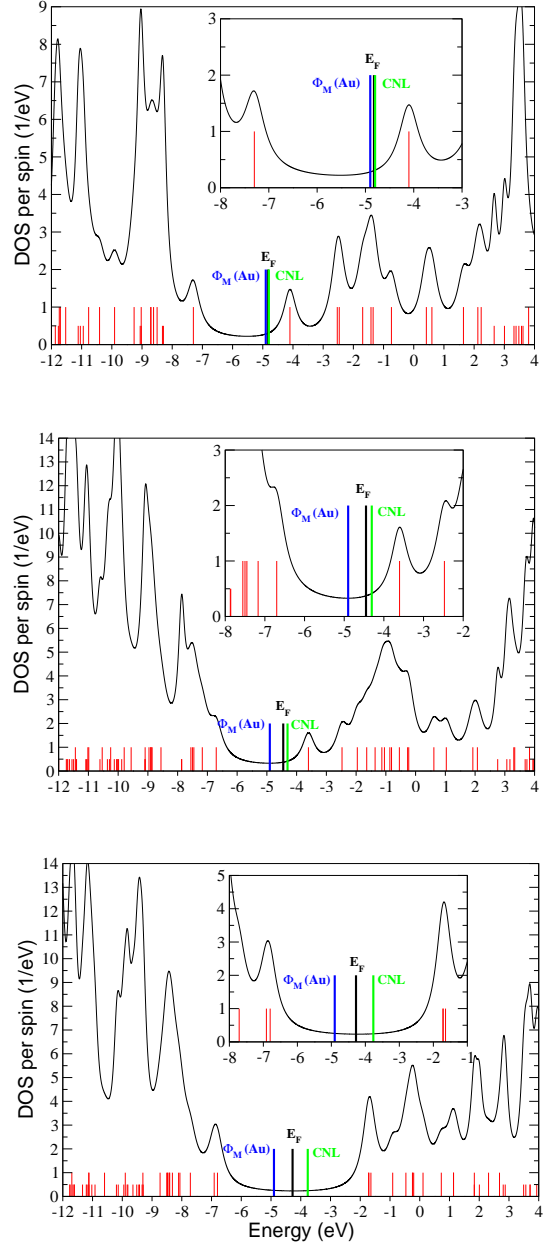

%\begin{figure}[t]
\begin{center}
\vspace{0.50cm}
\includegraphics[width=7cm, angle=0] {./PTCDA_fig4.eps}
\vspace{0.50cm}

\includegraphics[width=7cm, angle=0] {./PTCBI_fig4.eps}
\vspace{0.50cm}

\includegraphics[width=7cm, angle=0] {./CBP_fig4.eps}
\end{center}
\caption{(TOP to BOTTOM:) IDIS, CNL and interface Fermi level for PTCDA/Au, 
PTCBI/Au and CBP/Au. Long (short) bars correspond to the $\pi$ ($\sigma$) 
states neglecting the metal-molecule interaction. The insets show the region 
around $E_{F}$ in detail.
}
\label{Fig4}
\end{figure}

Figure~\ref{Fig4}  shows our results for PTCDA, PTCBI and CBP. In
these calculations $d=3.5$ \AA, $\Gamma_{i}^{\pi} \simeq 0.5$
eV and $\Gamma_{i}^{\sigma} \simeq 0.25$ eV. Note the different
broadening of the $\pi$ and $\sigma$ levels. These values
were obtained using equation~\ref{gammai6s} and the hopping
integrals of Figure~\ref{Fig3} for $d=3.5$ \AA. Note also the
uncertainty on the broadening, due to the uncertainty
in the metal-organic distance (see below).

Figure~\ref{Fig4} shows the different
molecular levels for each molecule, the $\pi$-($\sigma$-)states being drawn as long
(short) bars. For each molecule, the CNL is calculated by imposing charge 
neutrality conditions: the total
electronic charge below the CNL integrates to the number of occupied molecular 
states. Table~\ref{tab:1} shows the
calculated CNLs (measured from the ionization level). Also
included in this table are the transport gaps for the sake of
comparison. Note the similarity between PTCDA and PTCBI, with
the CNL level rather close to the affinity level. This is due to the
similarity of their energy spectra and transport gaps. In both cases the CNL 
is close to the affinity level because of the distribution of $\pi$-states 
around the energy gap. The large density of states below the ionization level
pushes the CNL upper in the gap. For CBP, the CNL is closer to the molecular 
midgap, due to the larger energy gap and
the more symmetric distribution of $\pi$-states around the HOMO and LUMO.

\begin{table}
\begin{tabular}
[c]{ccc}\hline\hline   & CNL (eV) & $E_{g}^{t} (eV)$ \\\hline
PTCDA & 2.5 & 3.2 \\ PTCBI & 2.4 & 3.1 \\ CBP & 3.0 & 5.1 \\
\hline
\end{tabular}
\caption{Charge Neutrality Levels (measured from the center of the HOMO) and 
peak-to-peak transport gaps, for the three organic materials.} \label{tab:1}
\end{table}

The interface slope parameter, $S = \frac {d E_{F}}{d \phi_{M}}$, is given by:
\begin{equation}\label{eqn:slope}
S = \frac{d E_{F} } {d \phi_{M}} =  \frac {1} {1 + 4 \pi e^{2}
D(E_{F}) \delta / A },
\end{equation}
where $D(E_{F})$ is the induced density of states at the Fermi
energy (twice the values shown in Figure~\ref{Fig4}), $d$ is the
metal-organic distance and $A$ is the area associated with one
organic molecule (see Figure~\ref{Fig1}; A=120 \AA$^2$, 191 \AA$^2$
and 251 \AA$^2$, for PTCDA, PTCBI and CBP respectively).
Table~\ref{tab:2} compares the experimental and calculated
values for the slope parameter. Note again the similarity
between PTCDA and PTCBI, and the larger value found for CBP. This
is due to the smaller $D(E_{F})$ and the larger area $A$. The different pinning
behaviour of the three interfaces is explained: the low value of $S$, as is the
case of PTCDA and PTCBI on Au(111), corresponds to a high pinning at the organic
CNL. For CBP/Au(111), on the other hand, the larger value of $S$ reflects the
higher ability of $E_{F}$ to move within the organic energy gap.

\begin{table}
\begin{tabular}
[c]{cccc}\hline\hline  & PTCDA/Au & PTCBI/Au & CBP/Au
\\\hline S (theory) & 0.2 & 0.2 & 0.5 \\ S (exp) & 0.0 & 0.0 & 0.6  \\ \hline
\end{tabular}
\caption{Calculated and experimental values for the interface slope 
parameter, S.} \label{tab:2}
\end{table}

This can be seen in the relation $CNL-E_F = S(CNL - \phi_M)$. Having calculated 
CNL and $S$, we obtain the interface Fermi level straightforwardly. The pinning 
at the interface reduces the initial difference $CNL - \phi_M$, to the 
injection barrier $E_F - \phi_M$.

Figure~\ref{Fig4} also shows the position of the Fermi level for 
interfaces with Au. The position, measured from the ionization level, is shown
for each molecular film in Table~\ref{tab:3}. The agreement between the
theoretical and experimental positions of $E_F$ and values of $S$ is remarkable, 
although a small
difference appears for the slope parameter. This presumably reflects the 
approximations introduced in our calculation. The
main source of inaccuracy comes from the assumed value of $d$,
which has an error bar of around $\pm 0.3$ \AA. This
inaccuracy is mainly reflected in the calculated values of
$D(E_{F})$, while the CNL is probably very insensitive to that
modification. To ascertain this point, we recalculated the CNL changing
$\Gamma_{i}^{\pi}$ and $\Gamma_{i}^{\sigma}$ by factors of up to 2,
and found that its value remains practically the same. This is not
the case for $D(E_{F})$, which changes by 50\% when $d$ changes $\pm 0.2$ \AA. 
Our results suggest that $d$ has probably
been overestimated for PTCDA and PTCBI by at least $0.3$ \AA ~in our
calculations.

\begin{table}
\begin{tabular}
[c]{cccc}\hline\hline  & PTCDA/Au & PTCBI/Au & CBP/Au
\\\hline $E_F$ (theory) (eV) & 2.5 & 2.3 & 2.5 \\ $E_F$ (exp) & 2.5 & 2.1 & 2.4  \\ \hline
\end{tabular}
\caption{Theoretical and experimental interface Fermi level positions, 
measured from the center of the HOMO.} \label{tab:3}
\end{table}

We stress, however, that our results for the interface Fermi level
show very good agreement with the experimental data.
Moreover, the main outcome of our analysis is to show that the
induced density of interface states is large enough to play a
crucial role in the formation of the metal/organic semiconductor barriers. This
allows us to conclude that the mechanism associated with the formation of these
interface barriers is the charge transfer between the two materials due to the
weak chemical interaction: this creates an electrostatic interface dipole 
which tends to align the
metal Fermi level and the organic CNL.

\acknowledgements
We gratefully acknowledge financial support by the Consejer\'ia de Educaci\'on
de la Comunidad de Madrid, the Spanish CICYT under project 
MAT 2001-0665, and the DIODE network (HPRN-CT-1999-00164). 
Support of this work by the National Science Foundation (DMR-0097133) and the
New Jersey Center for Organic Optoelectronics (AK) is also acknowledged.

\end{document}